\documentclass[pre,aps,showpacs,twocolumn]{revtex4}
\usepackage{epsfig}
\usepackage[T1]{fontenc}
\usepackage{ae,aecompl}
\begin{document}
\title{Morphologies and flow patterns in quenching of lamellar systems 
with shear}
\author{Aiguo Xu$^{(*)}$ and G. Gonnella}
\affiliation{Dipartimento di Fisica, Universit\`a di Bari\\
{\rm and} Istituto Nazionale di Fisica Nucleare, Sez. di Bari\\
{\rm and} INFM {\rm and}
 Center of Innovative Technologies for Signal Detection
and Processing (TIRES), \\
via Amendola 173, 70126 Bari, Italy}
\author{ A. Lamura}
\affiliation{Istituto Applicazioni Calcolo, CNR, Sezione di Bari,
via Amendola 122/D, 70126 Bari, Italy}
\date{\today}
\begin{abstract}
We study the behavior of a fluid quenched from the disordered into
the lamellar phase under the action of a shear flow. The dynamics
of the system is described by Navier-Stokes and
convection-diffusion equations with pressure tensor and chemical
potential derived by the Brazovskii free-energy. Our simulations
are based on a mixed numerical method with Lattice Boltzmann
equation and finite difference scheme for Navier-Stokes and order
parameter equations, respectively. We focus on cases where banded
flows are  observed with two different slopes for the component of
velocity in the direction of the applied flow. Close to the walls
the system reaches a lamellar order with very few defects and the
slope of the horizontal velocity is higher than the imposed shear
rate. In the middle of the system the local shear rate is lower
than the imposed one and the system looks as a mixture of tilted
lamellae, droplets and small elongated domains. We refer to this
as to a region with a Shear Induced Structures (SIS)
configuration. The local behavior of the stress shows that the
system with the coexisting lamellar and SIS regions is in
mechanical equilibrium. This phenomenon occurs, at fixed
viscosity, for shear rates under a certain threshold; when the
imposed shear rate is sufficiently large, lamellar order develops
in the whole system. Effects of different
 viscosities have been also considered: The SIS region is observed 
only at low enough
 viscosity. We compare the
above {\it scenario} with the usual one of shear banding. In
particular, we do not find evidence for  a plateau of the stress
at varying imposed shear rates  in the region with banded flow. We
interpret our results as due to a tendency of the lamellar system
to oppose to the presence of the applied flow.
\end{abstract}
\maketitle

\section{Introduction}

Complex fluids such as polymer solutions, liquid crystals, or
surfactant systems are characterized by the presence of organized
structures at  mesoscopic scales between   macroscopic and solvent
molecular lengths \cite{CE}. Under the action of external forcing,
 the coupling  between the mesoscopic structures and the local
velocity field makes the   flow properties of complex fluids
different from those of simple liquids \cite{larson}. For example,
when a  shear flow is applied to a simple fluid, after a short
initial transient, the linear relation between the  stress
$S$ and the shear rate
$\gamma$ is verified, $S= \eta \gamma$ with
$\eta$ being the fluid viscosity. In complex fluid an effective
viscosity can be defined by the same relation but its value is not
constant, depending on the strength of the applied flow. In
systems with interfaces, for examples, the effective viscosity
generally decreases (shear thinning) when the shear rate is
increased due to the alignment of interfaces with the flow.

In some cases the flow can induce new organization in the fluid
\cite{BRP94}. In lamellar phases, for instance, onion and
hexagonal phases have been observed not existing at rest
\cite{CBDCL,HBP,PAR}. Such shear induced structures (SIS) can
coexist in a range of applied shear rates with the structures
unmodified by the flow. Usually the SIS and the unmodified phase
have different viscosities so that they flow with different
profiles. This phenomenon, called shear banding, has been observed
in many complex fluids and also in lamellar systems
\cite{RND,SCM}. Its microscopic origin is not yet completely
understood and different explanations have been proposed depending
on the system \cite{Olmstedrev}.

The traditional theoretical description of shear banding and other
rheological behaviors in complex fluids is based on the assumption
of a local relation between the  stress and the  shear rate
\cite{EYB,olm99}. However, in this way, the role of  the
structures present in the fluid is not evident  and they   cannot
be directly related to the flow pattern. While a full description
of the system with all its molecular dynamical variables is not
possible, a description at mesoscopic level based on an order
parameter evolution equation can enlighten many issues concerning
the kinetics of shear banding, the morphology of the different
structures in the fluid and the evolution of the flow field
\cite{Onuki97,Yeomans}. Since the formation of SIS often occurs
during transient regimes  and since, also under stationary
conditions, the flow pattern is  not known {\it a priori}, one
understands that a description based on both
 Navier-Stokes and   order parameter equations is generally
required. The hydrodynamical description
is useful also for comparisons with experiments  where not always
the dynamical quantities are all easily accessible.

The aim of this paper is twofold. First, we want to show the
relevance of the full hydrodynamical description also in systems
with applied flow and see how a specific  problem can be
conveniently studied by our simulation methods. Then we hope to
clarify some aspects of  the kinetics of formation of lamellar
phases in cases when banded flows are observed \cite{olm03}. In
our model the lamellar properties are encoded in a free-energy
functional of an order parameter  representing the relative
concentrations of substances in the mixture. The order parameter
follows a convection-diffusion equation; from its instantaneous
configurations the stress can be calculated and inserted  in the
Navier-Stokes equation without assuming a further stress-shear
rate constitutive relation.

The free-energy functional considered in this paper was originally
introduced to study the effects of fluctuations in the
disordered-lamellar phase transition \cite{B75} and later used to
describe equilibrium properties of di-block copolymer systems
\cite{Lei80}. Solutions of copolymers, consisting of A-polymers
covalently bonded to B-polymers in pairs, can organize in striped
phases where A-rich and B-regions are separated by a stack of
lamellae \cite{Bat}. The order parameter in this case represents
the relative concentration of A and B substances. Our model is
also relevant for other systems with lamellar order. We mention
ternary mixtures where the surfactant form interfaces between oil
and water \cite{GS94},  dipolar \cite{SD} and supercooled liquids
\cite{KKZNT95}, chemically reactive binary mixtures \cite{GC}.

The dynamical  equations will be solved using a numerical scheme
based on a Lattice Boltzmann Method (LBM) for  the Navier-Stokes
equation and finite difference methods for the
convection-diffusion equation. LBM solves numerically a minimal
Boltzmann equation where the fluid can only move along the links
of a regular lattice with  dynamics consisting of a free-streaming
and a collision step \cite{lbe-1}. LBM has been largely applied to
the study of binary mixtures and complex flows
\cite{Yeomans,Swift96,yeo,noishear,lbe-2}. Rheological behavior of
liquid crystals under shear has been studied by LBM in
Ref.~\cite{DOY}.
 The  mixed method used in this paper \cite{luo,Xuepl,noidsfd}
allows a reduction in memory which is
convenient in large scale simulations. Moreover, spurious terms
appearing in Eq.~(\ref{motion2}) \cite{Swift96} are avoided and the
numerical efficiency is increased.

In our simulations we start from a disordered configuration and
consider the evolution of the system with parameters corresponding
to the lamellar phase. This correspond to a sudden quench;  shear
is applied for all the evolution  after the quench. Lamellae are
expected to align with the flow with order propagating  from the
moving walls \cite{tanaka}. One could think that after a
sufficient long time lamellae will be ordered uniformly in the
whole system. However, we will see  that the system can behave
differently. In some range of parameters we will find that a SIS
region develops in the middle of the system consisting of small
droplets and pieces of bent and rolled lamellae. This region
coexists with lamellar regions close to the walls and a banded
flow with different shear rates is observed. These  results will
be discussed in relation with the traditional {\it scenario} of
shear banding \cite{EYB,olm99,olmsted}.

The paper is organized as follows. In the next Section the
equilibrium model  and the dynamical equations will be first
introduced. Then the various components of the stress and the
structure factor will be defined. Finally, a short review of the
numerical scheme we use will be given. Section III contains our
results for the evolution of the system in a specific case where a
SIS region with banded flow is observed. The effects of changing
viscosity and shear rate on the properties of the  SIS region will
be shown in Section IV. In Section V we analyze the evolution of
global order in the system in the spirit of what usually done in
phase separation studies, considering the behavior of the first
momenta of the structure factor. A discussion with conclusions
will be given in Section VI.

\section{The model and the method}

Our simulations are based on a mixed numerical approach which
combines the lattice Boltzmann method with a finite difference
scheme \cite{Xuepl,noidsfd}. In this scheme the equilibrium properties
of the system can be controlled by introducing a free energy.

\subsection{Equilibrium properties and dynamical equations}

The equilibrium phase is described by a coarse grained free energy
that, for the specific problem we study here, is the following:
\begin{equation}
F \!=\! \int d {\bf r}\left[ \frac{1}{3} n \ln n +
\frac{a}{2}\varphi^2 +\frac{b}{4}\varphi^4+\frac{\kappa}{2}
(\nabla \varphi)^{2} + \frac{d}{2} (\nabla^2 \varphi)^2 \right]
\label{fren}
\end{equation}
where $n$ is the total density of the system and $\varphi$ is a
scalar order parameter representing the concentration difference
between the two components of the mixture. The term in $n$ gives
rise to a positive background pressure and does not affect the
phase behavior. The terms in $\varphi$ correspond to the
Brazovskii free energy \cite{B75}. We take $b,d > 0$ to ensure
stability. For $a>0$ the fluid is disordered; for  $a<0$ and
$\kappa > 0 $ two homogeneous phases with $\varphi = \pm
\sqrt{-a/b}$ coexist. A negative $\kappa$ favors the presence of
interfaces and a transition into a lamellar phase can occur.  In
single mode approximation,
 assuming a profile like $A \sin k_0 x$ for the direction transversal to the
lamellae,
 one finds   the transition ($|a|=b$) at  $ a \approx - 1.11 \kappa^2 /d $
 where  $ k_0 = \sqrt{ -\kappa /2d }$ and $A^2 = 4 (1 + \kappa^2/4db)/3$
\cite{Xuepl}.

The evolution of the system is described by a set of two coupled
partial differential equations: The Navier-Stokes and the
convection-diffusion equations. The fluid local velocity ${\bf u}$
obeys, by assuming incompressibility  (${\bf \nabla} \cdot {\bf u}
= 0$), the Navier-Stokes equation which reads as
\begin{equation}
n \Big ( \partial_t  u_{\alpha}+
 {\bf u} \cdot  {\bf \nabla} u_{\alpha} \Big )
= -  \partial_{\beta} P_{\alpha \beta}^{th} + n \nu \nabla^2
u_{\alpha} \;\;\;\; , \label{motion1}
\end{equation}
where $\nu$ is the kinematic viscosity. $P_{\alpha \beta}^{th}$ is
the thermodynamic pressure tensor which can be calculated from the
free energy functional (\ref{fren}) as
\begin{equation}
P_{\alpha\beta}^{th} = \Big \{ n \frac{\delta F}{\delta n} +
\varphi \frac{\delta F}{\delta \varphi} - f(n, \varphi) \Big \}
\delta_{\alpha \beta} + D_{\alpha \beta}(\varphi) \label{pres}
\end{equation}
where $f(n, \varphi)$ is the free-energy density and a symmetric
tensor $D_{\alpha \beta}(\varphi)$ has to be added to ensure
 that the condition of mechanical equilibrium
$\partial_{\alpha} P_{\alpha\beta}^{th}=0$ is satisfied
\cite{evans}. The complete expression of the pressure tensor is
\cite{yeo}
\begin{equation}
P_{\alpha\beta}^{th} = p \delta_{\alpha\beta} +
P_{\alpha\beta}^{chem} \label{presstot}
\end{equation}
with
\begin{eqnarray}
&&p =   \frac{1}{3} n +\frac{a}{2} \varphi^2 + \frac{3b}{4}
\varphi^4 -\kappa \big [ \varphi ( \nabla^2 \varphi ) +\frac{1}{2}
( \nabla \varphi )^2 \big ] \nonumber \\
&&+ d \big [ \varphi ( \nabla^2 )^2
\varphi +\frac{1}{2} ( \nabla^2 \varphi )^2 + \partial_{\gamma}
\varphi \partial_{\gamma} ( \nabla^2 \varphi ) \big ]
\label{pdiag}
\end{eqnarray}
and
\begin{equation}
P_{\alpha\beta}^{chem} = \kappa \partial_{\alpha} \varphi
\partial_{\beta} \varphi
- d \big [
\partial_{\alpha} \varphi \partial_{\beta} ( \nabla^2 \varphi )
+  \partial_{\beta} \varphi \partial_{\alpha} ( \nabla^2 \varphi )
\big ] \label{poff}
\end{equation}
Moreover, we shear the system by moving the upper and lower walls
with velocities
\begin{equation}
{\bf v}_{u,l} = \pm \gamma \frac{L}{2} {\bf i} , \label{flow}
\end{equation}
respectively, where $\gamma$ is the shear rate, $L$ is the width
of the system  and ${\bf i}$ is a unit vector along the $x$-axis,
which is usually denoted as the flow direction. The presence of
the moving walls greatly affects, as we will see, the
fluid velocity ${\bf u}$ and the behavior of the  order parameter
$\varphi$.

The evolution of $\varphi$ is described by the
convection-diffusion equation:
\begin{equation}
\partial_t \varphi + {\bf \nabla} \cdot (\varphi {\bf u})
= \Gamma \nabla^2 \mu \label{motion2}
\end{equation}
where
\begin{equation}
\mu = \frac {\delta F} {\delta \varphi}= a \varphi + b \varphi^3 -
\kappa \nabla^2 \varphi + d (\nabla^2)^2 \varphi \label{chem}
\end{equation}
is the chemical potential
\cite{footnote},
 $\Gamma$ is a mobility coefficient, and the laplacian
in the r.h.s. of (\ref{motion2}) guarantees the conservation of
$\varphi$.

\subsection{The shear stress}

The presence of shear strongly influences the morphology of the
system and  its rheological properties. In particular, we will
consider the effects on shear stress.

The total stress is (see the r.h.s. of Eq.~(\ref{motion1}))
\begin{equation}
S_{\alpha\beta} = - P_{\alpha\beta}^{th}
+ n \nu \left ( \partial_{\beta} u_{\alpha} + \partial_{\alpha} u_{\beta}
\right ).
\label{totstress}
\end{equation}
The total shear stress, which is related to the
off-diagonal part of the total stress (\ref{totstress})
\cite{larson}, is
\begin{equation}
S_{xy} = - P_{xy}^{chem} + n \nu \left ( \partial_{y} u_{x}
+ \partial_{x} u_{y} \right )
\label{totstress1}
\end{equation}
$S_{xy}$ is the sum of the time reversible
shear stress $S_{xy}^{chem} = - P_{xy}^{chem}$ and of
a dissipative
hydrodynamic contribution $S_{xy}^{hydr} = n \nu (\partial_{y}
u_{x}+\partial_{x} u_{y})$. The expression (\ref{totstress1})
depends on  local coordinates and may vary from point to point
especially when the system is not homogeneous.

A quantity of relevant experimental interest is the integral
\begin{equation}
S = \int d {\bf r} S_{xy} = S^{chem}+S^{hydr} =\int d {\bf r}
S_{xy}^{chem}+ \int d {\bf r} S_{xy}^{hydr} \label{totstress2}
\end{equation}
By using Eq.~(\ref{poff}) it can be shown that $S^{chem}$ can be
also calculated in the reciprocal space as
\begin{equation}
S^{chem}=\int d {\bf r} S_{xy}^{chem} = \int \frac{d {\bf k}}{(2
\pi)^d} k_x k_y (\kappa + 2 d k^2) C({\bf k},t) ,
\label{totstress3}
\end{equation}
where $C({\bf k},t)$ is the structure factor
\begin{equation}
C({\bf k},t)= \langle \varphi({\bf k}, t)\varphi(-{\bf k},
t)\rangle ,  \qquad
\end{equation}
$\varphi({\bf k})$ is the Fourier transform of the order
parameter $\varphi$ and $ \langle \cdot \rangle $ is the average
over different histories.

\subsection{The numerical scheme}

The Eqs.~(\ref{motion1}-\ref{motion2}) are numerically solved in
2D by using a mixed approach. We use the lattice Boltzmann method
for Eq.~(\ref{motion1}) and a finite difference scheme for
Eq.~(\ref{motion2}). Such an approach has been already adopted in
the case of thermal lattice Boltzmann models for a single fluid
and for multiphase flows \cite{luo}. In that case it is the
temperature equation to be solved by finite differences. By using
this approach we are able to avoid the spurious terms
in the convection-diffusion equation (\ref{motion2}) which
come into play when standard LBM for binary mixtures is used
\cite{Swift96}, though LBM may realize boundary conditions easily
and give better numerical stability. The present method allows to
reduce required memory.

A set of distribution functions $f_{i}({\bf r}, t)$ is defined on
each lattice site ${\bf r}$ at each time $t$. Each function is
associated to a lattice speed vector ${\bf e}_i$ with $e_i/c =$
$(\pm 1, 0), (0, \pm 1), (\pm 1, \pm 1), (0, 0)$, where $c=\Delta
x / \Delta t$, $\Delta t$ is the time step, and $\Delta x$ is the
lattice constant.

They evolve according to a single relaxation-time Boltzmann
equation \cite{bhat,lbe-1}:
\begin{equation}
f_{i}({\bf r}+{\bf e}_{i}\Delta t, t+\Delta t)-f_{i}({\bf r},
t)= -\frac{1}{\tau}[f_{i}({\bf r},t)-f_{i}^{eq}({\bf r},t)]
\label{dist1}
\end{equation}
where $\tau$ is a relaxation parameter and $f_{i}^{eq}({\bf r},
t)$ are local equilibrium distribution functions. They
are related to the total density  $n$ and
to the fluid momentum $n {\bf u}$ through
\begin{equation}
n=\sum_{i}f_{i}, \;\;\;\; n {\bf u}= \sum_{i}f_{i}{\bf e}_{i}.
\label{vel}
\end{equation}

These quantities are locally conserved in any collision process
and, therefore, we require that the local equilibrium distribution
functions fulfil the equations
\begin{equation}
\sum_i f_i^{eq}=n, \;\;\;\;  \sum_i f_i^{eq} {\bf e}_i=n {\bf u} .
\end{equation}
Following Ref.~\cite{Swift96}, the higher moments of the local
equilibrium distribution functions are defined so that the
Navier-Stokes equation can be obtained and the equilibrium
thermodynamic properties of the system can be controlled:
\begin{equation}
\sum_{i}f_{i}^{eq}e_{i\alpha}e_{i\beta}=c^2 P_{\alpha\beta}^{th}+n
u_{\alpha} u_{\beta} .
\end{equation}
The local equilibrium distribution functions can be expressed as
 an expansion at the second order in the
velocity ${\bf u}$ \cite{Swift96}. The expression of the
coefficients of the equilibrium distribution functions can be
found in Ref.~\cite{noi}.

\begin{figure*}
\epsfig{file=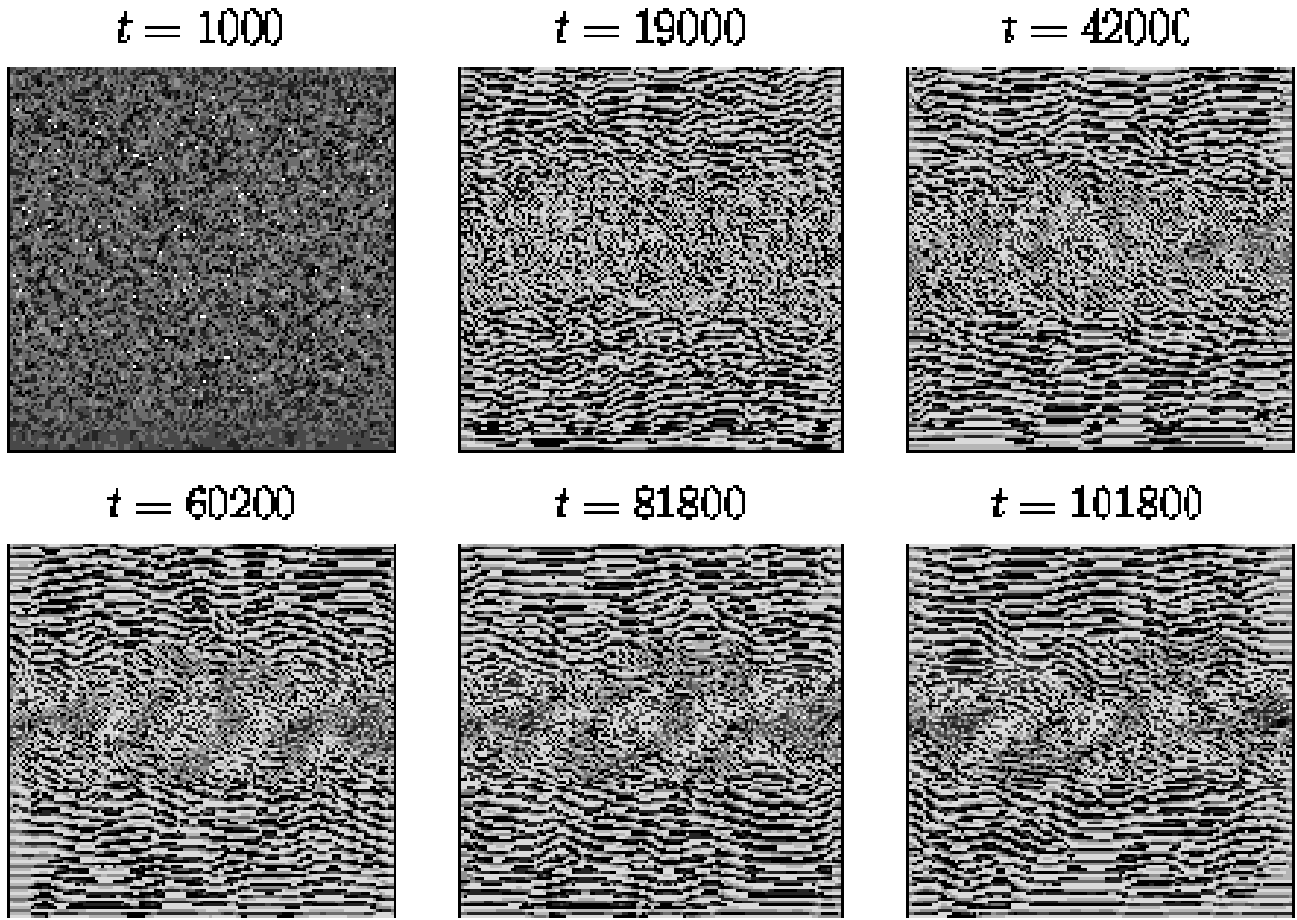,bbllx=67 pt,bblly= 243 pt,bburx=529 pt,
bbury= 577pt,width=17.9 cm,clip=}
\caption{Configurations of the order parameter $\varphi$ on the
whole $1024 \times 1024$ lattice for the case with $\gamma=5
\times 10^{-5}$ and $\nu=4.1667$.} \label{fig1}
\end{figure*}

The above described lattice Boltzmann scheme simulates at second
order in $\Delta t$ the continuity and the incompressible
Navier-Stokes equations (\ref{motion1}) with the kinematic
viscosity $\nu$ given by
\begin{equation}
\nu=\Delta t \frac{c^2}{3} (\tau-\frac{1}{2}) . \label{visc}
\end{equation}
It appears that the relaxation parameter $\tau$ can be used to
tune independently the viscosity.

Equation (\ref{motion2}) is numerically integrated by using an
explicit Euler scheme on a square lattice with spacing $\Delta x$,
the same as for LBM. The spatial derivatives are approximated by
discrete expressions which are second order in $\Delta
x$. The time step is $\Delta t' = \Delta t / m$ with $m=5$. This
choice was motivated by the observation of a better numerical
stability of the code.

To enforce the flow (\ref{flow}) we assume periodic boundary
conditions (BC) along the flow direction and we place walls at the
upper and lower rows of the lattice moving them at a constant
velocity along the $x$ direction avoiding slip velocity
\cite{noishear}. The velocity ${\bf u}$ obtained from
Eq.~(\ref{vel}) goes inside the convection-diffusion equation
(\ref{motion2}). For the order parameter $\varphi$ we adopt
Lees-Edwards BC along the $y$ direction \cite{lees}. This means
that $\varphi(x,-L/2) = \varphi(x+(v_u-v_l)\Delta t',L/2)$.
The algorithm implementing
the previous numerical scheme has been described in
Ref.~\cite{noidsfd}.

All the simulations in the following have been run by using the parameters
$-a = b = 2 \times 10^{-4}$, $\kappa = -6 \times 10^{-4}$,
$d = 7.6 \times 10^{-4}$, and $\Gamma = 25$. The system size was $L=1024$ and
space and time steps were set to $\Delta x =1$ and $\Delta t = 0.2$,
respectively.
At the beginning of each run
the values of $\varphi$ are randomly taken in the range $[-0.1,0.1]$,
the distribution functions $f_i$ are set so that $n=1$,
and the velocities ${\bf u}$
are computed from Eq.~(\ref{vel}).
We verified that by vertically shifting the reference frame,
the fluid velocity profile is accordingly shifted. Therefore we
used the choice of placing the walls symmetrically located at
$y = \pm L/2$ to have fluid almost at rest in the middle of the system.

\section{Kinetics of SIS formation}

In this Section we will show the  evolution of the system described
by Eqs.~(\ref{motion1},\ref{motion2}) for a typical case  where shear
induced structures appear. Shear rate and viscosity are fixed to
$\gamma= 5 \times 10 ^{-5}$ and
$\nu=4.1667$.  All the quantities here and
in the following are measured in units of the
space step $\Delta x$ and the time step $\Delta t$.
For these values,  the relaxation time of the linear
shear velocity profile in a simple fluid would be of the order of
$t=6000$ \cite{footnote2}.

Figure \ref{fig1} shows the configurations of $\varphi$ at
successive times. At $t=1000$ after the quench, the order
parameter has locally reached one of the minima of the polynomial
part of the free-energy,  represented by black and white in the
figure.  Lamellae are ordered only on small scales and in most of
the system the effects of shear flow are not observable. Ordered
structures appear only very close to the walls. At the next time
$t=19000$, lamellar order has developed into the system, but the
the middle region is still almost isotropic apparently scarcely
influenced by the flow. After this time the region of lamellae
aligned with the flow does not increase the extension towards  the
central part of the system where  morphology evolves  in a
different way. At $t= 42000$ the middle region mostly consists of
lamellae oriented at about $45^{o}$ with respect to the direction
of the flow. Few domains are broken into droplets and small
pieces of lamellae.
The interface between the  central and the two external lamellar regions
is not very sharp and shows some undulations.

The  further evolution of the middle region can be seen at
$t=60200$. We will call this region as a SIS region or SIS phase.
Many ruptures have occurred in the central network with the
consequent formation of more droplets and worm-like domains. Not
relevant changes can be observed at $t=81800$ and $t=101800$.
Droplets, once formed, are quite stable. We checked the existence
of the SIS region until time $t=200000$. We repeated this
numerical experiment starting from 5 different initial
configurations and obtaining very similar results for the
different histories.

\begin{figure}
\epsfig{file=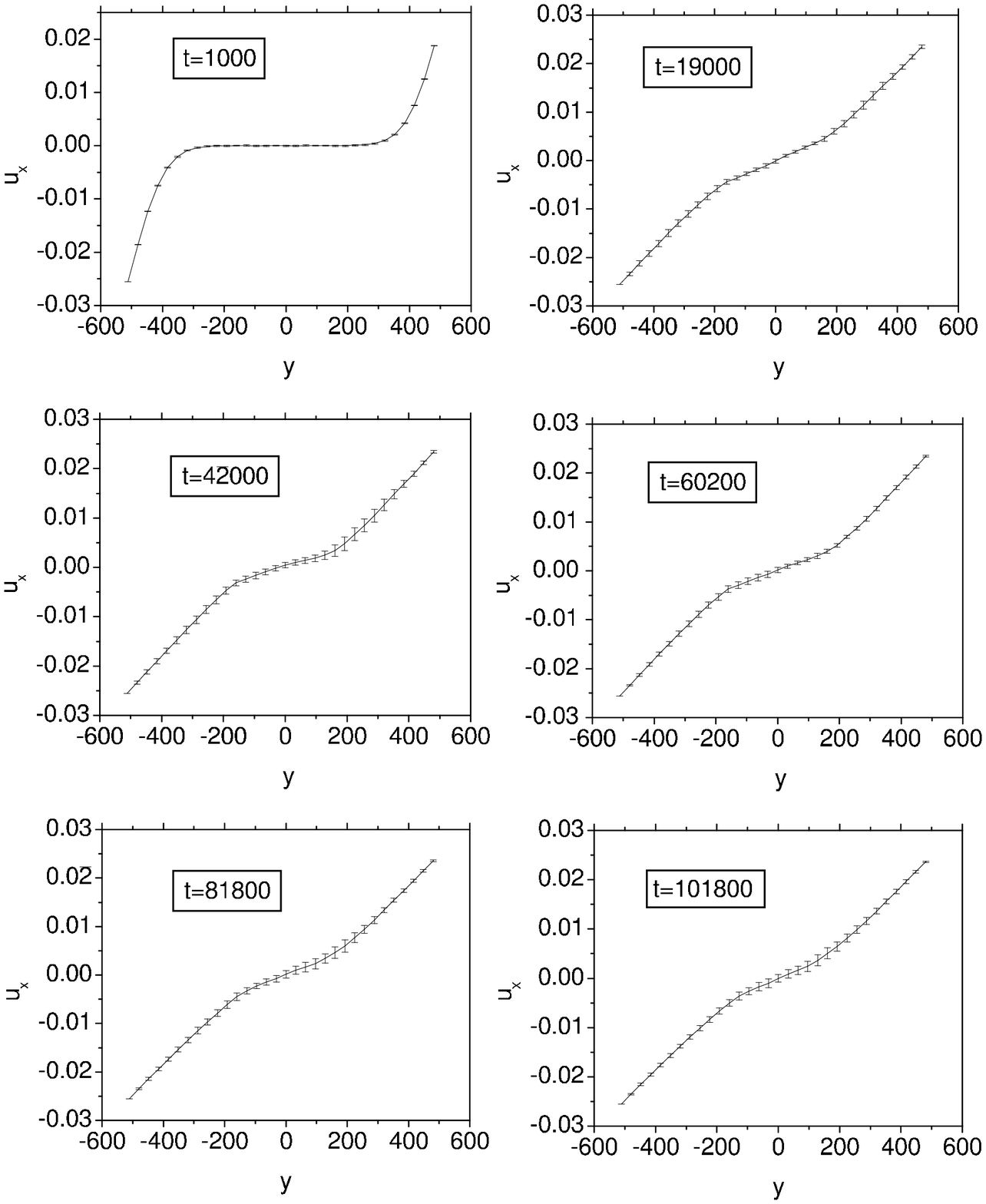,bbllx=28 pt,bblly= 73 pt,
bburx=566 pt,bbury= 730pt,width=8.5 cm,clip=}
\caption{Plots of the $x$-averaged horizontal velocities across the system
at different times for the case
with $\gamma=5 \times 10^{-5}$ and $\nu=4.1667$.}
\label{fig2}
\end{figure}

The morphological evolution of the system  has to be examined also
in relation with the behavior of the velocity profile. In
Fig.~\ref{fig2}  the $x$-averaged horizontal velocities $u_x$ are
plotted at the same times of Fig.~\ref{fig1} a functions of $y$.
The variance corresponding to each average is also shown; it is
generally quite small indicating an uniform behavior of the system
in the flow direction. At the beginning ($t=1000$) the shear
profile is different from  zero only in a region of about $150$
lattice sizes close to the walls; this explains the isotropy of the
configuration observed  almost everywhere  at this time in
Fig.~\ref{fig1}. At later times the $u_x$-profile becomes
characterized by  two slopes found in correspondence  of the
regions with lamellar and SIS phases. The profile remains almost
stationary from  $t \sim 40000$; the local shear rates in the
lamellar and SIS regions are respectively higher and lower than
the imposed value.

\begin{figure}
\epsfig{file=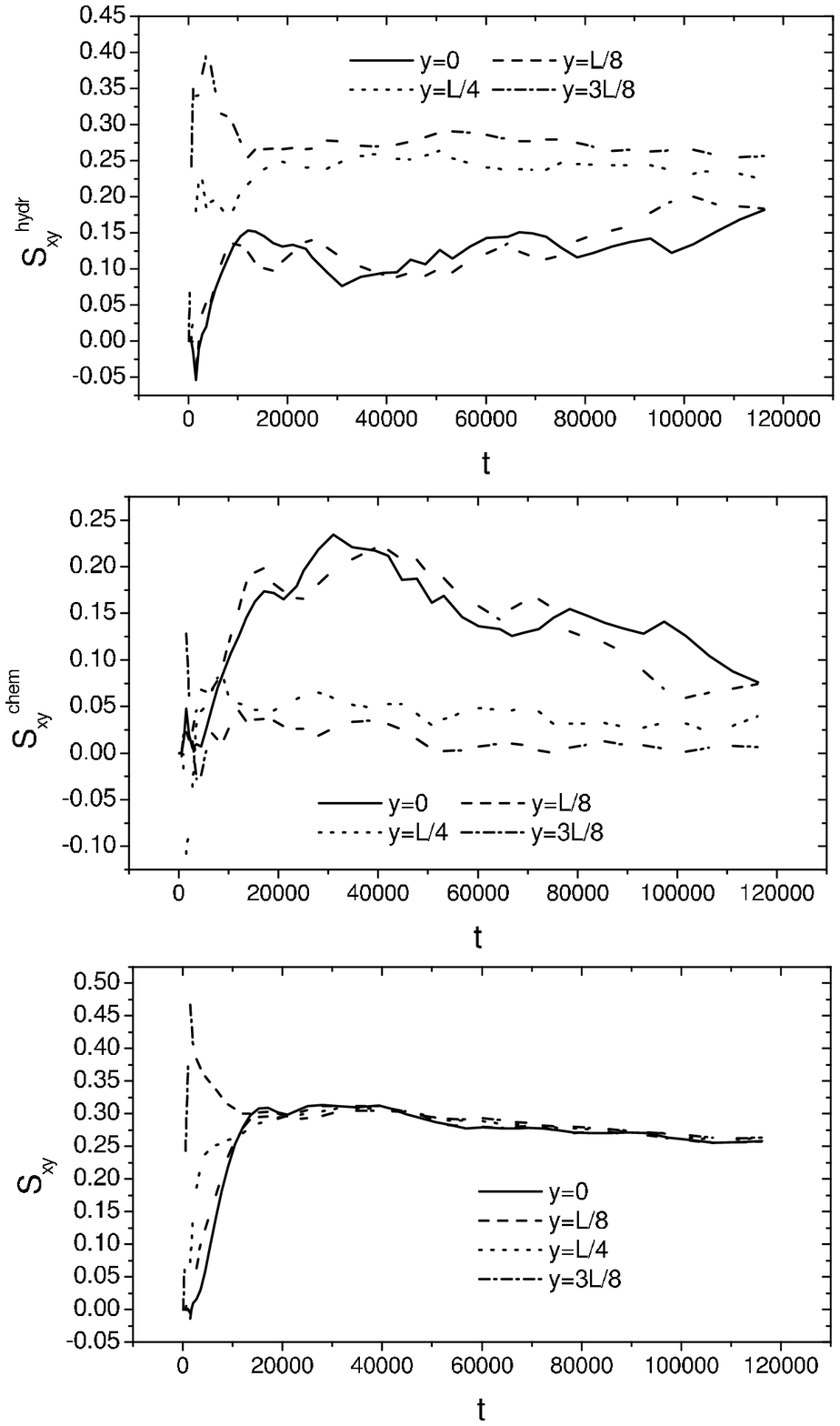,bbllx=97 pt,bblly= 133 pt,bburx=450 pt,bbury=
730pt,width=8.5 cm,clip=} \caption{Plots of the different
components of stress, summed over the $x$-axis, as functions of
time at different positions across the system for the case with
$\gamma=5 \times 10^{-5}$ and $\nu=4.1667$.} \label{fig3}
\end{figure}

Mechanical properties are described by the behavior  of the two
stress $S_{xy}^{chem}$ and $S_{xy}^{hydr}$. The structures present
in the system locally determine the value of the chemical part of
the stress. $S_{xy}^{chem}$   is larger in presence of defects or
configurations which are not minima of the free-energy while it is
close to zero for a well ordered lamellar configuration. This can
be seen in the central part of Fig.~\ref{fig3} where we compare
the chemical stress at different distances from the walls. In the
middle of the system ($y=0$) the $x$-averaged stress reaches a
maximum in correspondence of the formation of the SIS  and then
remains different from zero. Fluctuations correspond to the
evolution of structures present in the SIS phase. The behavior of
the chemical stress at $y=L/8$, just inside the SIS region, is
similar. On the other hand, in the region with lamellar order, at
$y=L/4$ and $y=3L/8$, after a maximum at initial times, the stress
relaxes to a very small value. This maximum occurs  before than
lamellae become aligned with the flow. This behavior is analogous
 to that observed in phase separation  of binary mixtures under
shear where the excess viscosity reaches a maximum before than
domains orientate with  interfaces in the direction of the flow
\cite{giapp}.

The hydrodynamical stress, also shown in Fig.~\ref{fig3}, has an
opposite behavior: It is larger where shear rate is larger
($y=L/4$ and $y=3L/8$). Closer to the walls ($y=3L/8$) it relaxes
to a constant value after an initial maximum corresponding to the
fact  that the shear rate close to the walls is higher at initial
times. The bottom part of Fig.~\ref{fig3} shows that the system is
in mechanical equilibrium from $t\sim 20000$ onwards when the
total stress has become the same at different distances from the
walls.

\begin{figure}
\epsfig{file=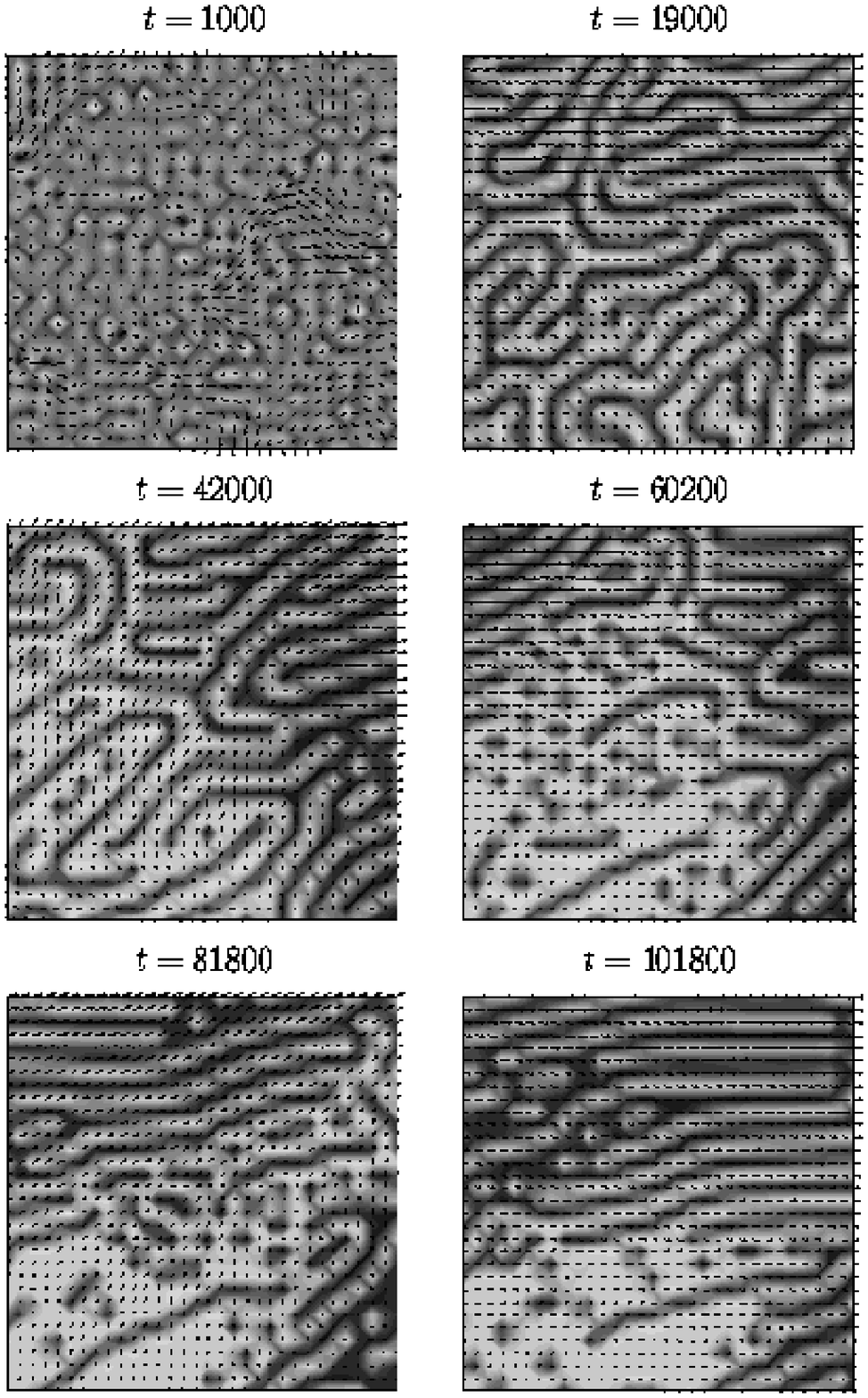,bbllx=73 pt,bblly= 68 pt,bburx=503 pt,bbury=
760 pt,width=8.6 cm,clip=} \caption{Configurations of the order
parameter $\varphi$ on a central portion of size $128 \times 128$
of the whole lattice for the case with $\gamma=5 \times 10^{-5}$
and $\nu=4.1667$. Arrows represent fluid velocity vectors ${\bf
u}$ and are plotted every four lattice sites.} \label{fig4}
\end{figure}

Finally, in Fig.~\ref{fig4},  the global patterns of the velocity
field are shown for the same times of Fig.~\ref{fig1}. We consider
a central
region of the system  which includes a portion of the SIS phase
and the interface with the lamellar region. At $t=1000$ there is
no evidence of the imposed flow in this part of the system;
local flows, as usually in phase segregation, originate from  the
defects. Effects of the imposed flow becomes evident at $t=19000$.
Moving inside the SIS phase, the magnitude of the velocity vector
becomes smaller with horizontal and vertical components
comparable. Vortex structures, not present in the lamellar phase,
can be also observed. In later pictures one observes  that  the
SIS-lamellar interface structure  evolves with time together with
the local flow patterns.  Velocity in the SIS phase always remains
small and  a jump in the  velocity magnitude can be seen moving
through the SIS-lamellar interface at all the  late times
considered.

\section{Properties of  SIS phase}

\begin{figure}
\epsfig{file=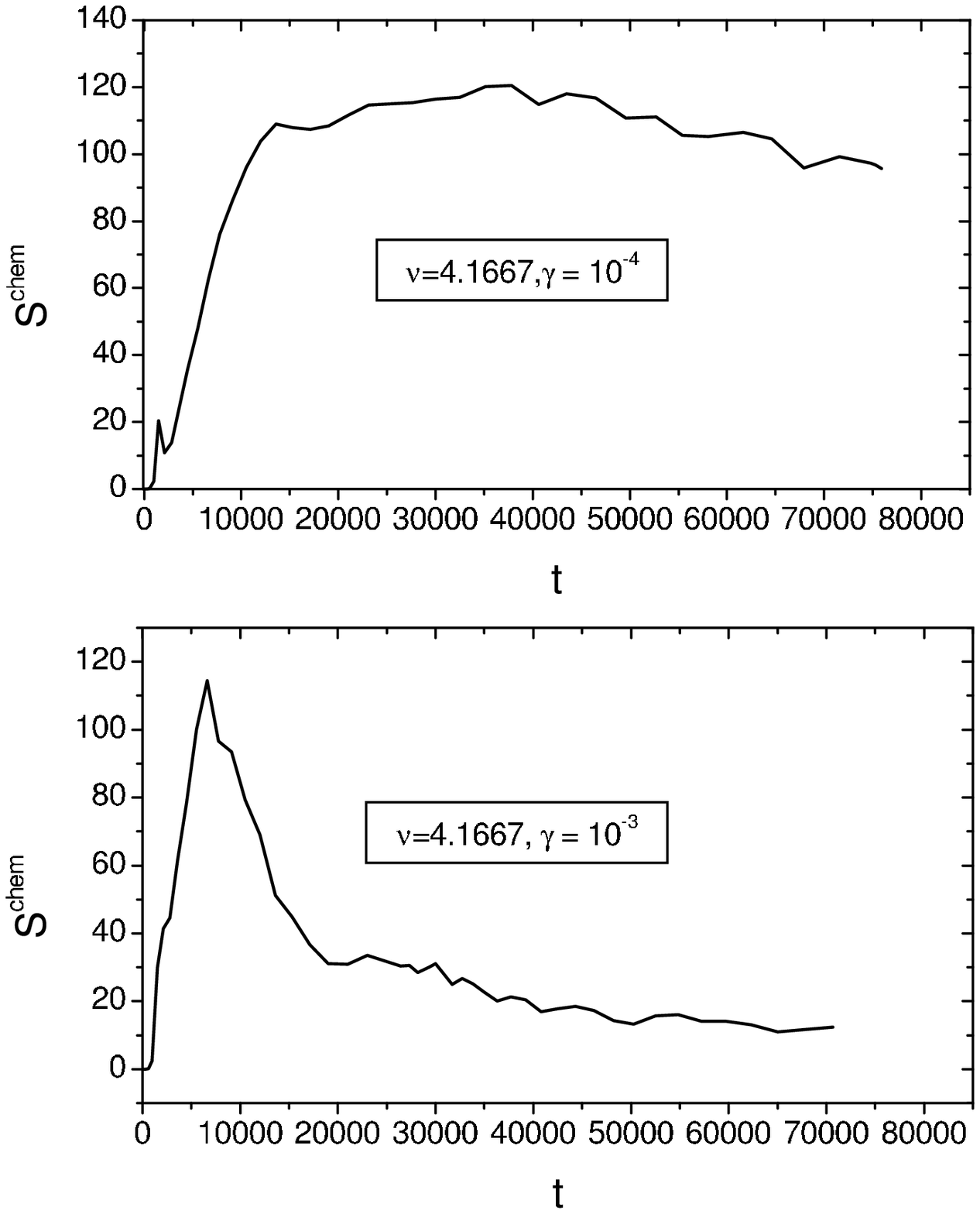,bbllx=80 pt,bblly=200 pt, bburx=500 pt,bbury=
730pt,width=8.5 cm,clip=} \caption{Plots of the integral chemical
stress  (Eq.~(\ref{totstress3})) as a function of time for two
cases with different shear rates.} \label{fig5}
\end{figure}

The behavior described in the previous Section is found, at a
given  viscosity, for shear rates under a certain threshold.
Velocity, stress and other properties, in cases when a SIS region
is observed, have been checked to remain stable for long times and
a linear velocity profile is never reached. Figure \ref{fig5}
shows the behavior of the chemical part of the stress for  two
values of shear rate smaller and larger than the threshold that,
for the case $\nu=4.1667$, is $ 2.5 \times 10^{-4} \lesssim
\gamma_c \lesssim 5 \times 10^{-4} $.

\begin{figure*}
\epsfig{file=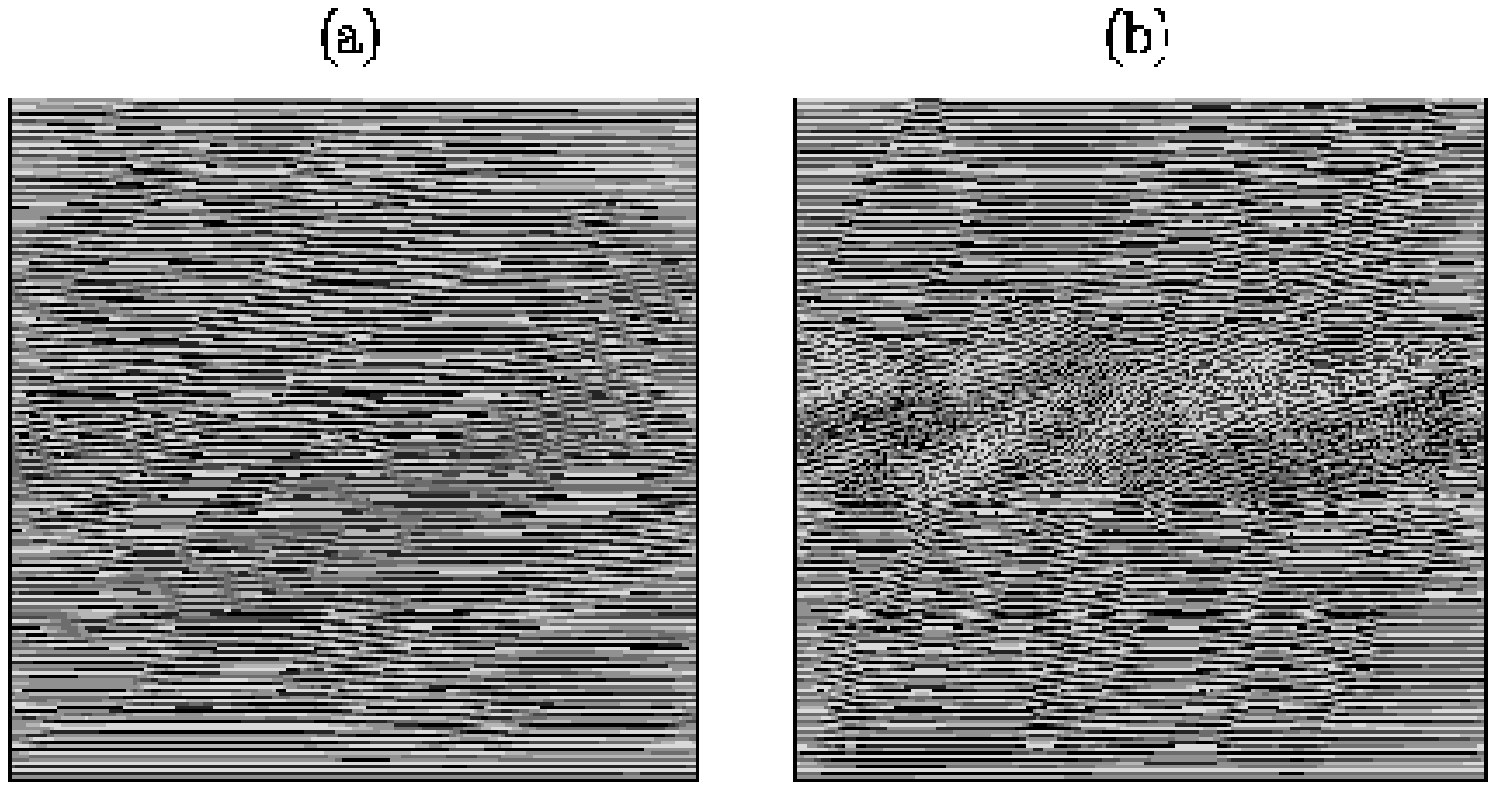,bbllx=72 pt,bblly= 293 pt,bburx=522 pt,bbury=
530pt,width=12 cm,clip=} \caption{Late time configurations of the
order parameter $\varphi$ on the whole $1024 \times 1024$ lattice
with $\nu=4.1667$ in the cases with $\gamma= 10^{-3}$ at time
$t=26400$ (a) and $\gamma=10^{-4}$ at time $t=73900$ (b).}
\label{fig6}
\end{figure*}

At  $\gamma = 10^{-4}$ the chemical stress remains almost
stationary after the formation of the SIS region and the
morphological evolution is similar to that shown in
Figs.~\ref{fig1}-\ref{fig2}. On the other hand, at $\gamma =
10^{-3}$, after a maximum, the shear stress  decreases to small
values while the horizontal velocity tends to a profile with
constant slope. In this case  the SIS region does not
 form in the middle of the system and ruptures do not occur at all in the
lamellar network.
 Late-time configurations for
the two cases are shown in Fig.~\ref{fig6}. For $\gamma = 10^{-3}$
one sees a well-ordered lamellar state with only few local
defects.

For the case $\nu=4.1667$ we run various simulations with
different values of $\gamma \leq \gamma_c$ also for comparing our
results for the SIS phase with the usual {\it scenario} of shear
banding \cite{Olmstedrev,EYB,olmsted,hamley}.
In this {\it scenario} bands of phases with different
structure and flow properties can coexist  only  at a critical
value of the stress $S^{\star}$. In experiments with imposed shear
rate this leads to a stress plateau on the flow curve (stress {\it
versus} imposed shear rate) at $S^{\star}$. On the plateau, the
shear rates of the different bands do not change by varying the
imposed flow which only determines the relative spatial extension
of the bands.

\begin{figure}
\epsfig{file=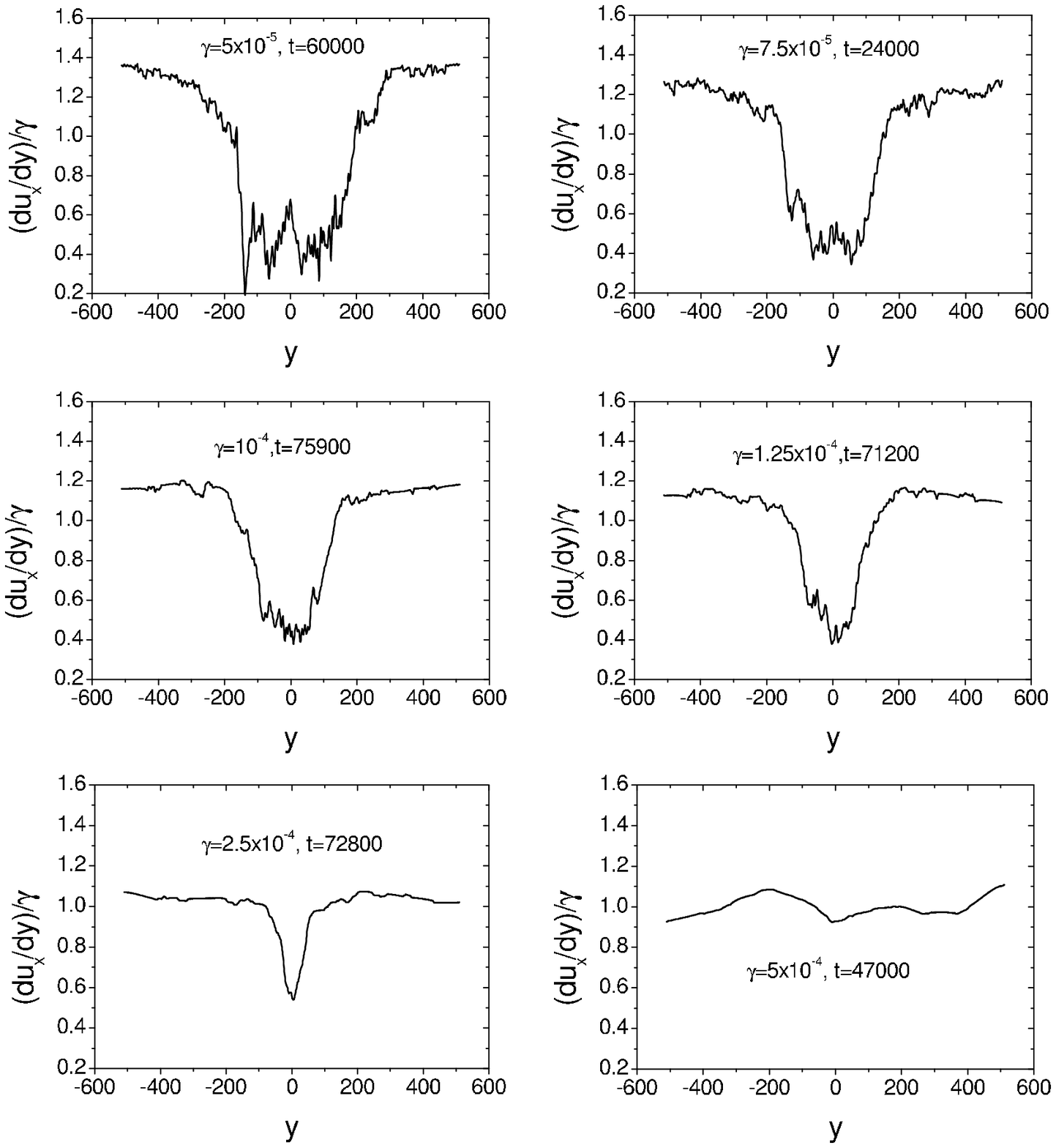,bbllx=20 pt,bblly= 170 pt,bburx=530 pt,bbury=
700pt,width=8.5 cm,clip=} \caption{Plots of the $x$-averaged
measured shear rate across the system for different imposed shear
rates $\gamma$ with $\nu=4.1667$. Values are  normalized to the
imposed shear rate $\gamma$.} \label{fig7}
\end{figure}

In Fig.~\ref{fig7} we show the  shear rate profiles for different
values of $\gamma < \gamma_c$ and for $ \gamma = 5 \times 10^{-4}$
 at times, for  $\gamma < \gamma_c$,  when  the SIS phase is well
formed. A central region with smaller shear rate can be clearly
identified in all cases, except that for $\gamma = 5 \times
10^{-4}$. The width of this region, the one with  SIS morphology,
decreases when the imposed sear rate is increased, as usually in
systems with shear banding. Close to the walls the shear rate is
larger than the imposed value with an almost constant value in
correspondence of the region with well aligned lamellae. However,
differently from other systems with shear banding, the values of
shear rates in the lamellar and in the SIS regions change with the
imposed flow. The constraint that the integral of the horizontal
velocity in the vertical direction gives  the value imposed on the
boundaries is always verified. An interface of finite width
between the lamellar and the SIS region can be seen in all cases,
confirming the relevance of the arguments discussed in
Ref.~\cite{olmsted}.

\begin{figure}
\epsfig{file=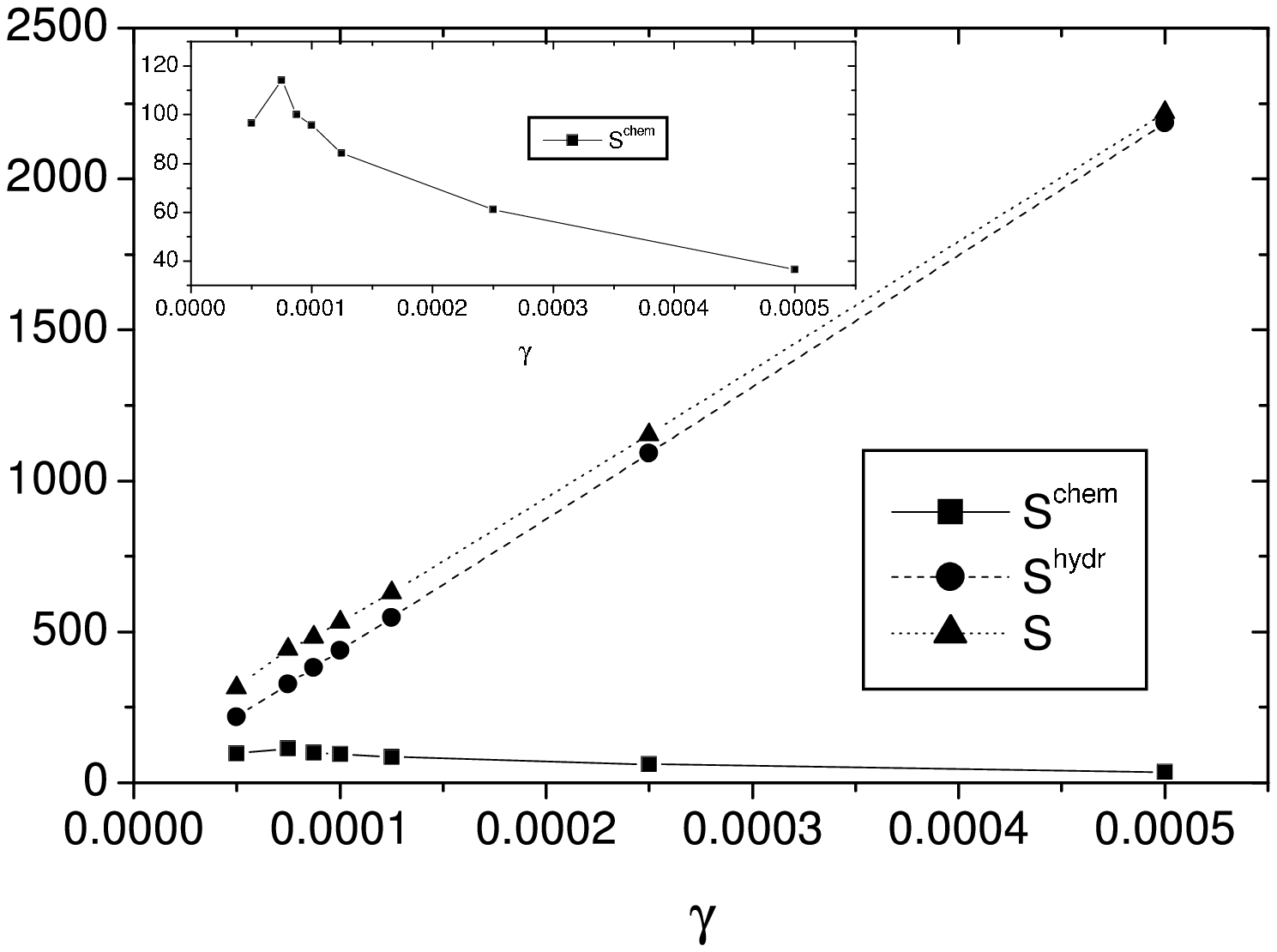,bbllx=45 pt,bblly= 330 pt,bburx=513 pt,bbury=
650 pt,width=8.5 cm,clip=}
\caption{Plots of the different
components of stress summed over the whole system at long times as
functions of the imposed shear rate $\gamma$ for the case with
$\nu=4.1667$. In the inset the chemical stress is plotted on a
magnified scale for a better view.}
\label{fig8}
\end{figure}

The behavior of the hydrodynamical, chemical and total stress is
shown in Fig.~\ref{fig8} as a function of the imposed shear rate.
The total stress does not exhibit a flat regime in the interval
where the SIS phase exists and changes in the  behavior are not
observed when  $\gamma$ becomes greater than $ \gamma_c$. As
expected, the chemical stress decreases with the reduction of the
width  of  the SIS region.

\begin{figure*}
\epsfig{file=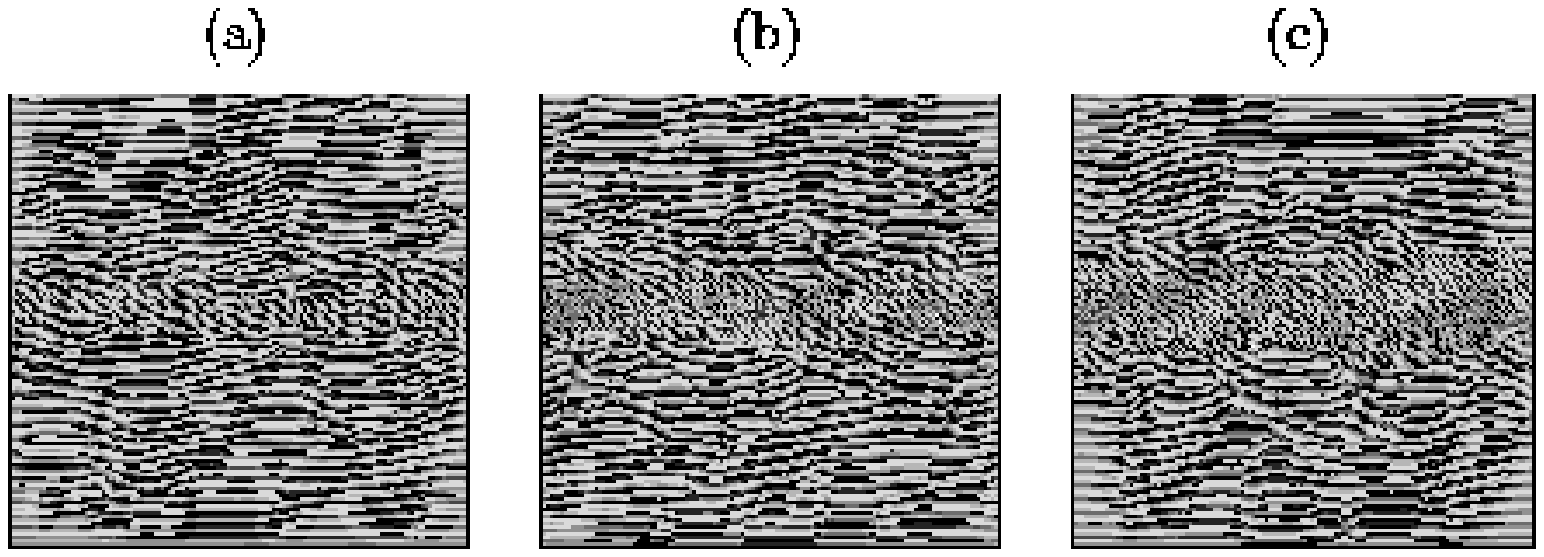,bbllx=66 pt,bblly= 325 pt,bburx=531 pt,bbury=
495pt,width=17.0 cm,clip=} \caption{Configurations of the order
parameter $\varphi$ on the whole $1024 \times 1024$ lattice for
the cases with $\gamma=10^{-4}$ and $\nu=15.883$ (a), $7.5$ (b),
$4.1667$ (c).} \label{fig9}
\end{figure*}

\begin{figure}
\epsfig{file=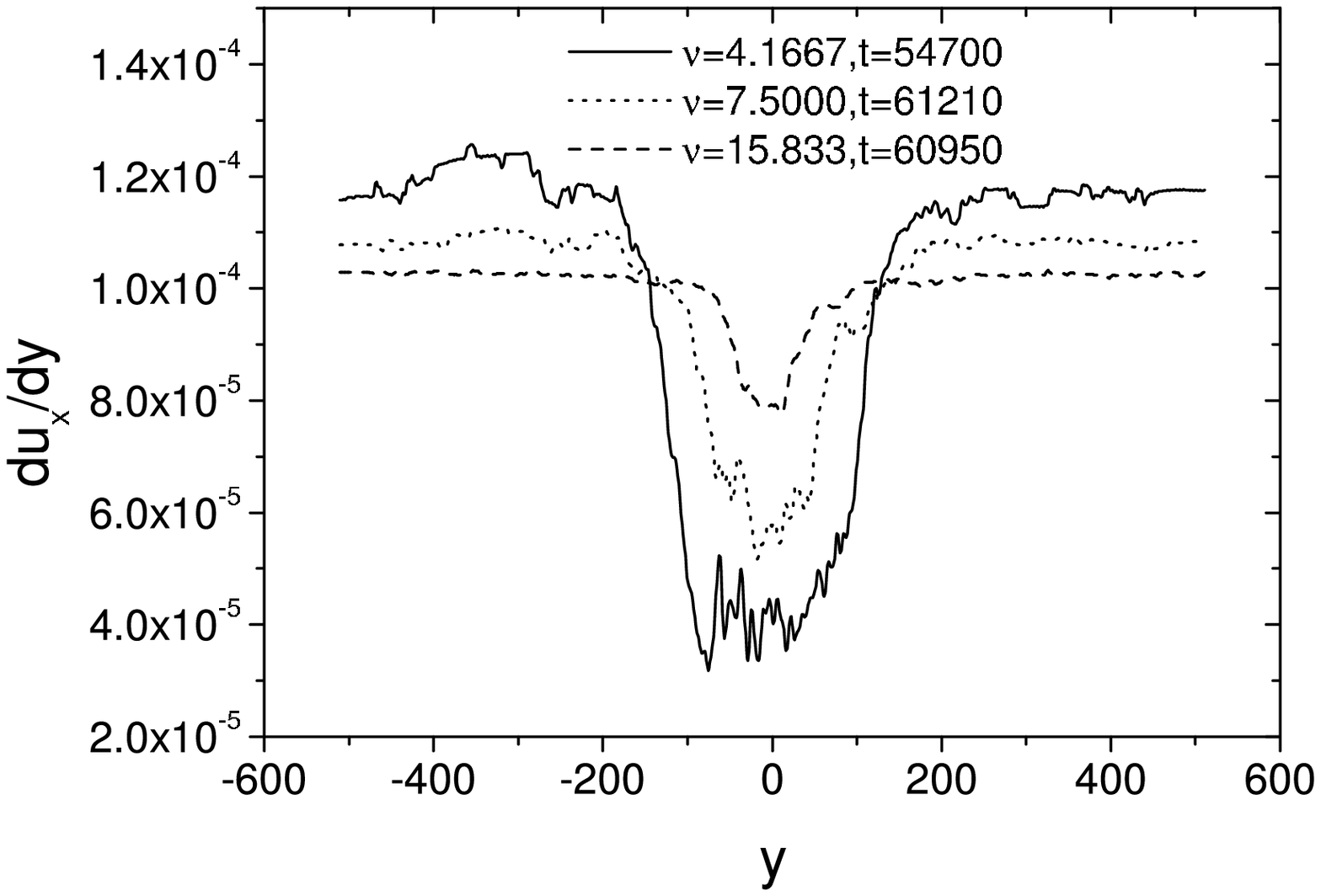,bbllx=30 pt,bblly= 330 pt,bburx=520
pt,bbury= 680pt,width=8.5 cm,clip=} \caption{Plots of the
$x$-averaged measured shear rate across the system for different
viscosities $\nu$ and $\gamma=10^{-4}$.} \label{fig10}
\end{figure}

Finally, we examine the effects of viscosity showing results at
fixed shear rate for times after   the SIS formation. The main
result is that at higher viscosities the width of the SIS region
becomes narrower, as shown in Fig.~\ref{fig9}.
 This behavior is confirmed  from Fig.~\ref{fig10}
where  shear rate  profiles are plotted.

We have tried to interpret these results considering possible
relations between  cases with a SIS region and  characteristic
time scales of the imposed flow. The shear rate naturally
introduces a typical time $\tau_{\gamma} = \gamma^{-1}$ when shear
effects are expected to become relevant \cite{Onuki97}. On the
other hand, the propagation of the horizontal velocity from the
moving walls requires a relaxation time which, in simple fluids,
is inversely proportional to the viscosity \cite{footnote2}. This
time is also relevant for the effectiveness of the imposed flow in
the middle of the system. We did not find a quantitative
explanation of the phenomena before described in terms of these
time scales, probably, also because the relaxation time of a
simple fluid is not appropriate for the system we are studying.
However, our results suggest the following general qualitative
observation. If shear effects arrive early enough in the middle of
the system, due to a large shear rate or to a high viscosity, the
flow is able to penetrate in this region and lamellae align with
it. Otherwise, at lower shear rates or lower viscosity, a SIS
region can be found. We will return on this point in the
discussion of the last Section.

\section{Domain Growth properties}

In cases without imposed flows,  the kinetics of formation  of
ordered phases after a quench is a  process which is well-known
for simple systems like binary mixtures, and  still under
investigations for lamellar systems. In binary mixtures dynamical
scaling occurs with  domains growing isotropically with power-law
behavior \cite{B94}. In lamellar systems the formation of extended
defects or grain boundaries between domains of differently
oriented lamellae characterizes the late time evolution; other
defects like dislocations and disclinations are also present
\cite{Har00,BV02,Xuepl}. Typical lengths can be defined and their
evolution follow power-law or slower asymptotic growth depending
on the steepness of the potential energy and on the depth of the
quench. In cases with power-law behavior the length corresponding
to  the inverse of the structure factor at half height grows with
an exponent ranging in the interval $0.2 < z < 0.33$
\cite{EVG92,BV01,CB98,Har00,YS02,QM02,Xuepl}.

\begin{figure}
\epsfig{file=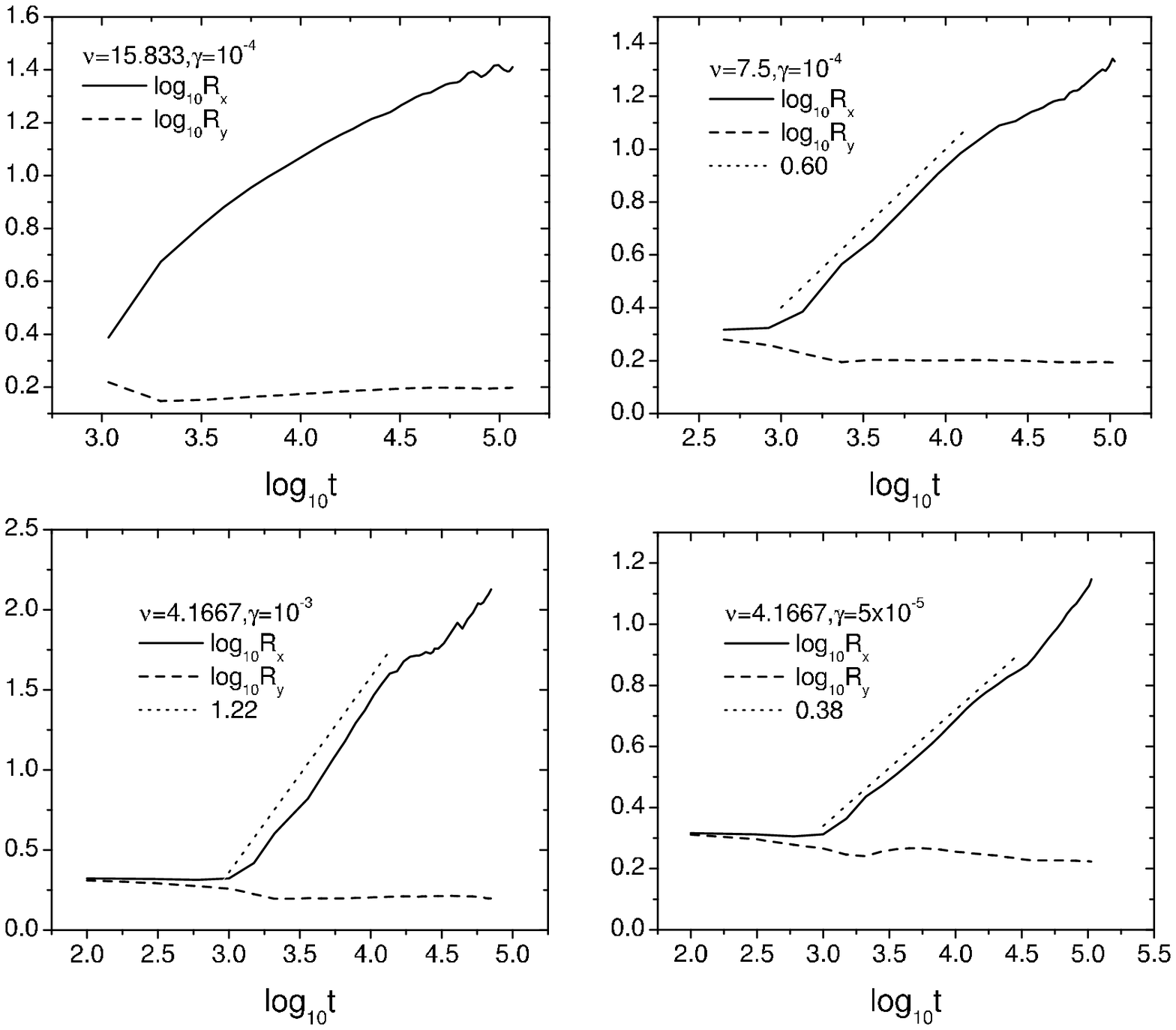,bbllx=50 pt,bblly= 280
pt,bburx=550 pt,bbury= 720
pt,width=8.5cm,clip=}
\caption{Plots of the domains sizes $R_x$ and $R_y$ as functions of time
for different values of viscosities and imposed shear rates. See the
single frames for parameters details and slopes of straight lines.}
\label{fig11}
\end{figure}

In our case with  shear the presence of banded structures makes
problematic the description of ordering in terms of a single
length, as we have seen in the previous sections. However, it
could still be  interesting to consider the behavior of averaged
lengths intended as macroscopic indicators of the degree of order
in the system. We define
\begin{equation}
R_{\alpha} = \frac{\int d {\bf k} C({\bf k},t)}{\int d{\bf k} |k_{\alpha}|
C({\bf k},t)}  \;\;\;\;\;\;\; \alpha = x , y
\end{equation}
and show in Fig.~\ref{fig11} the evolution of $R_x, R_y$ for
different viscosities and shear rates. In all cases $R_y$ relaxes
to the equilibrium value, while $R_x$
 grows  with exponents $z_x$ varying with the viscosity and the shear rate as
reported in the figure. In particular, keeping the viscosity $\nu$
fixed, we found
 that the value of $z_x$ increases by
increasing the shear rate $\gamma$. In the case discussed in
Section III with $\gamma = 5 \times 10^{-5}$ it results that $z_x
= 0.38 $ for times $t \leq 40000$ corresponding to SIS formation;
the later growth is characterized by the elimination of local
defects in the lamellar region. Finite size effects may affect the
behavior of $R_x$ at very late times. The case not shown in
Fig.~\ref{fig11} with $\gamma = 10^{-4}$ and the same viscosity
$\nu = 4.1667$ is similar. The time interval corresponding to the
SIS formation ($ t \leq 20000$) is characterized by a growth
compatible with exponent $z_x=0.52$ while the growth rate changes
later. By increasing further the shear rate ($\gamma = 10^{-3}$),
banded configurations do not form (see
Figs.~\ref{fig5}-\ref{fig6}) and the whole system reaches a
lamellar order with  exponent $z_x \sim 1.2$. The behavior at
$\gamma = 10^{-4}$ and $\nu = 7.5$ is qualitatively similar to
those already discussed; the difference is in the value of the
growth rate  which is now $z_x = 0.6$ during the SIS formation. We
observe that this value is greater than in the case with the lower
viscosity $\nu = 4.1667$. Finally, for the highest value of
viscosity shown in Fig.~\ref{fig11}, a power law  appears not
appropriate for  describing the behavior of $R_x$. We can conclude
that the growth of lamellar order as measured by $R_x$ is faster
for parameters corresponding to a smaller extension of the SIS
region and a larger lamellar phase.

\section{Discussion and Conclusions}

In this paper we have shown results from  two-dimensional
simulation results for  a fluid mixture quenched  from a
disordered configuration into a state with lamellar order. The
fluid, described by Navier-Stokes and convection-diffusion
equations, is subject during all the evolution to the action of a
shear flow imposed by the walls of the system. The ordering
process of a lamellar system, in absence of flow,  would be
characterized by the presence of local and extended defects that
make slow, sometimes freezing,  the evolution of the system. When
shear is applied, defects tend to be eliminated and lamellae would
be  expected to align in a preferred direction forming
well-ordered macroscopic domains.

Our results actually show that  different evolutions with more
complex morphologies are also possible under shear. For small
enough shear rates and viscosities we found that the flow
stabilizes itself with a horizontal velocity profile characterized
by two different slopes. Close to the walls the shear rate is
higher than the imposed one and well-ordered lamellae can be
observed. In the central part of the system, that is reached later
by the flow, the shear rate is smaller than the imposed one and
the morphology of domains is characterized by the presence of
small droplets and pieces of bent or rolled lamellae never aligned
with the flow. We referred to this region as to a SIS phase.

Shear banding phenomena occur in many complex fluids. Similarly to
what generally found, in our case  the width of the SIS phase
decreases by increasing the value of the imposed shear rate. In
other important aspects, however, our results differ from the
usual picture of shear banding. In the range of shear rates with
coexisting  lamellar and SIS regions, we do not observe a plateau
of the total stress at varying shear intensity. Moreover, in this
range, the values of local shear rates corresponding to the two
phases depend on the imposed flow.

At the moment, we are not aware of experiments with the same
scenario as that described in this paper. However, even if our
model can be appropriately used  for studying copolymer systems in
the weak segregation limit \cite{Lei80},  the comparison with
experiments is limited  from the fact that our simulations are
two-dimensional while in three-dimensional systems more complex
geometries can occur \cite{fre,hamley}, with possible different
rheological behaviors and flow patterns.  Actually, the  purpose
of this work  was more generic. We wanted  to analyze general
features of the  formation of banded flows in systems with
lamellar order  considering the full dynamical problem for the
velocity and the concentration fields.

Our results of Section III and IV show that when the applied flow
is weak enough  or propagates from the walls sufficiently slowly
(at lower viscosities), it is not able to penetrate with the same
shear rate in all the system. In this case the central region,
with its intertwined tangled lamellae,
opposes to the presence of the flow and tends to keep its
morphology. The  region tends to behave as an almost frozen
network; the  shear mainly acts stressing this network and causing
ruptures with the production of  droplets and small pieces of
lamellae. Even if the shear rate in the SIS region is quite small
but not zero, the behavior of our system resembles that of yield
stress fluids where no flowing steady states exist for stresses
under a certain threshold \cite{pabl}. For example, in soft glass
systems, coexistence between not flowing "pasty" states and
sheared fluid regions has been observed \cite{v3b}. In relation
with this, it  is interesting to observe  that an equilibrium
glass transition has been found also in systems with lamellar
order \cite{SW00}. We think that in our case the competition
between the intrinsic slow dynamics \cite{Xuepl} and the
acceleration induced by the external flow is responsible for the
peculiar shear banding phenomena we have  shown.

Finally, we observe  that our numerical methods have been proven
to be quite convenient  for simulating
fluids with complex order under driving forces. These methods can
be implemented also with different geometries and in the more
realistic three-dimensional case. We hope that our two-dimensional
results can be useful for the comprehension of shear banding and
stimulating for new experiments.

\begin{acknowledgments}
We warmly thank  P. L. Maffettone and J. M. Yeomans for
helpful discussions. We acknowledge
 support by  MIUR (PRIN-2004).
\end{acknowledgments}

$^{(*)}$ Present address: Division of Physics and Astronomy,
Yoshida-south Campus, Kyoto University, Sakyo-ku, Kyoto, 606-8501
Japan.


\begin{thebibliography}{999}
\bibitem{footnote}
In terms of the chemical potentials $\mu_1$ and $\mu_2$ of the two
components of a mixture one has $2 \mu = \mu_1 - \mu_2$.
\bibitem{footnote2}
The relaxation time of a linear shear profile in a simple fluid is
given by $L^2 / (4 \pi^2 \nu)$.
\end{thebibliography}
\end{document}